\documentclass[11pt,twoside]{book} 
\usepackage{asp2008s}
\usepackage{times}
\usepackage{lscape}
\usepackage{epsf}

\usepackage{graphicx}
\usepackage{amssymb}
\usepackage{longtable}
\usepackage[figuresright]{rotating}
\usepackage{multirow}

\newcommand{\simless}{\mathbin{\lower 3pt\hbox {$\rlap{\raise 5pt\hbox{$\char'074$}}\mathchar"7218$}}}

\mainmatter

\newlength{\deftabcolsep}
\begin{document}

\title{The Ara OB 1a Association}   
\author{Scott J. Wolk}   
\affil{Harvard--Smithsonian Center for Astrophysics, 60 Garden Street, Cambridge, MA 02138; USA }    
\author{Fernando Comer\'on}   
\affil{European Southern Observatory,Karl-Schwarzschild-Strasse 2, 85748 Garching, Germany}    
\author{Tyler Bourke}   
\affil{Harvard--Smithsonian Center for Astrophysics, 60 Garden Street, Cambridge, MA 02138; USA}    

\begin{abstract} 
The Ara OB1a association is one of the closest sites where triggered
star formation is visible for multiple generations of massive
stars. At about 1.3 kpc distance, it contains complex environments
including cleared young clusters, embedded infrared clusters, CO
clouds with no evidence of star formation, and clouds with evidence of
ongoing star formation. In this review we discuss the research on this
region spanning the last half-century.  It has been proposed that the
current configuration is the result of an expanding wave of neutral
gas set in motion between 10--40 million years ago in combination with
photoionization from the current epoch.

\end{abstract}



\section{Historical Summary (1960--1985)}

\begin{figure}[!ht]
\begin{center}
\includegraphics[scale = 0.5, angle = 0]{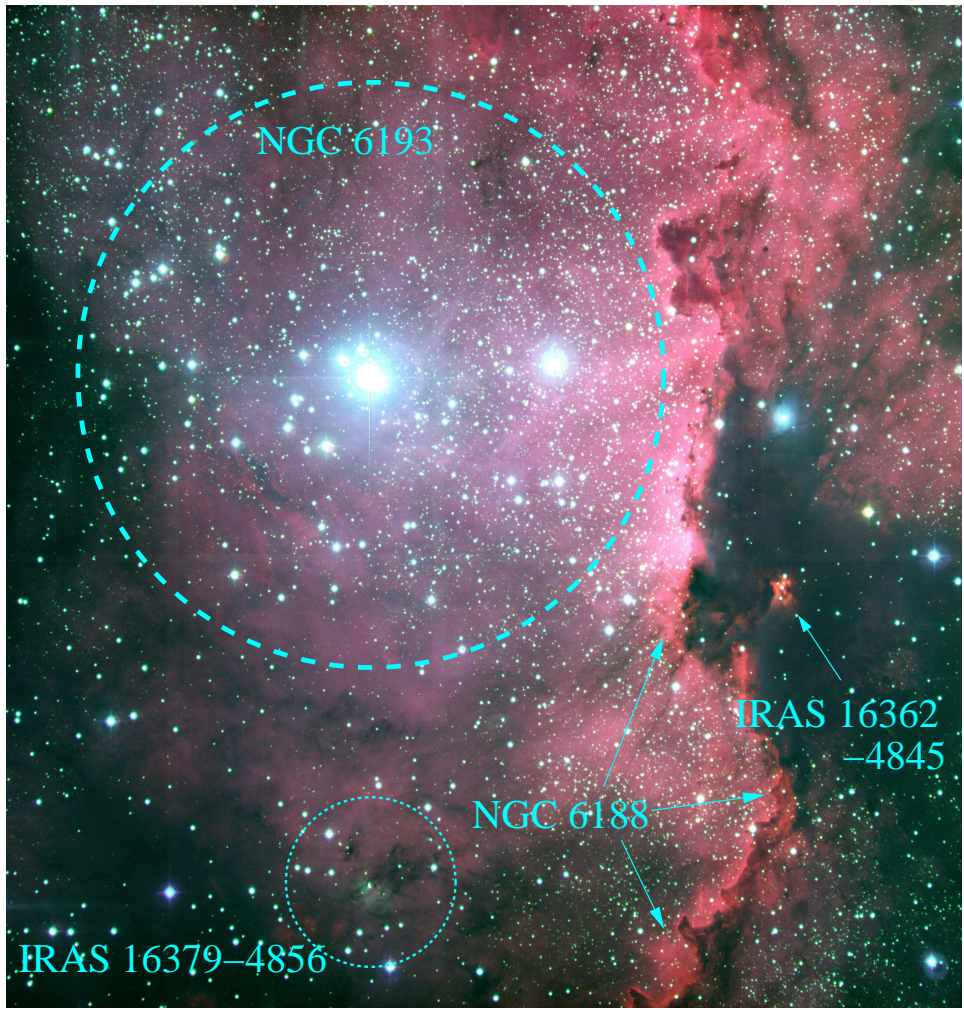}
\end{center}
\caption{Salient features of the most active portion of Ara OB1a. This
  image is a B,V H$\alpha$ (blue, green and red respectively) from the
  wide field imager on the MPI-ESO 2.2m telescope. The field of view
  is 32\arcmin x 32\arcmin\ or about 12 pc on a side.} \label{ESO}
\end{figure}

Ara OB1a, while somewhat enigmatic, may be one of the best examples of
triggered star formation in the local Galaxy.  The triggering in the
most active portion is easily imagined from images of the $<$ 3 Myr
NGC~6193/RCW~108-IR complex such as Figure~\ref{ESO}.  However the
full association may be as much as 50 Myr in age and cover several
degrees on the sky.  Ara OB1 was first studied in detail by Whiteoak
(1963). He used photographic photometry and objective prism
spectroscopy to identify about 35 O and B star members.  Whiteoak's
work was a follow up to an H$\alpha$ survey by Rodgers, Campbell \&
Whiteoak (RCW; 1960), which cataloged 181 H$\alpha$ emission regions
in the southern sky, including RCW 108.  NGC~6193 is an open cluster,
discovered by James Dunlop in 1826, that is dominated by a pair of O
stars, HD 150135 and HD 150136.  These are the brightest optically
revealed O stars in the association and are thought to be responsible
for ionizing the bright rim of emission to the west (NGC 6188,
discovered by John Herschel in 1836) which separates NGC 6193 from RCW
108--IR. The youth of the region is clear in sky survey plates and
Figure~\ref{ESO} which show concurrent regions of ionized gas and dust
lanes. Whiteoak noted two additional clusters in the region, NGC~6204
-- about 2 degrees to the northeast -- and NGC~6167 -- about 1 degree to
the southwest.

Ara OB1a is a compact association covering about 1 sq. degree around a
central cluster -- NGC 6193. It is generally thought to be about 1.3
kpc away and can be equated with RCW~108.  There is some confusion in
the literature as to what is actually meant by RCW~108. The original
definition of Rodgers, Campbell \& Whiteoak (1960) refers to all the
region where H$\alpha$ nebulosity is detected, for which they give a
size of 210\arcmin x 120\arcmin centered at ({\em l,b} = 336.49,
--1.48). Straw et al. (1987) used RCW~108--IR to refer to the embedded
IR cluster about 15\arcmin\ to the west of the O stars in NGC~6193 and
which is identified with IRAS~16362--4845.  The confusion arises when
RCW~108-IR is abbreviated by dropping the "--IR".  For the remainder of
this paper we will refer to the embedded cluster as RCW~108--IR or
IRAS~16362--4845.

Moffat \& Vogt (1973) measured photometry for 13 stars within
4\arcmin\ of the center of NGC 6193 and found E$_{B-V}$= 0.4 and a
distance of 1360 pc.  Herbst \& Havlen (1977) describe Ara OB1
as having a ``diamond ring'' appearance, with RCW 108--IR as the
``diamond'' and a thin circular dust ring making up almost half of the
``ring''.  They performed photoelectric and photographic photometry on
702 stars in the region.  Herbst (1974, 1975) identified parts of this
region as an ``R association'' as there were three early type stars
associated with reflection nebulosity.  These stars may mark a separate site of
star formation within Ara OB1a.  We will discuss this further in
Section 6.

Mid--infrared MSX \& IRAS maps of the region
show several condensations.
The brightest is coincident with RCW~108--IR
 and two apparently related peaks  IRAS 16379--4856
 and IRAS~16348--4849.  
One of the earliest radio studies of the region was in the survey by
Shaver \& Goss (1970).  They made a 5 GHz and 408 MHz survey of over
250 Galactic radio source including RCW 108. RCW 108 was remarkable
for having one of the smallest emission regions -- unresolved at
3\arcmin\ and having a relatively high density and temperature of the
electron population.


In addition to Ara OB1a, Whiteoak (1963) identified a background O
star cluster coincident with NGC~6193 but with a different distance
modulus.  While the foreground Ara OB1a has a distance modulus of
10.5$\pm 0.6$ the second group of about 13 O and B stars (Ara OB1b)
has a distance modulus of 12.7 $\pm 0.5$ or about 3500 pc. The central
O stars of the two associations are offset by about 2$^{\rm o}$ along
the Galactic plane (Humphreys 1978), so this is a concern for
membership determination.

\section{NGC 6193}

NGC~6193 is the open cluster to the east of the bright emission rim
NGC~6188.  In this region, Herbst \& Havlen (1977) measured over 700
stars photometrically. Of these, 59 had photometric distances and
reddening consistent with cluster membership.  This appears to be
complete to earlier than A0. This survey covered an area about
40\arcmin\ on a side.  They found a slightly steeper than usual
reddening law.  Their best-fit photometric distance was 1320$\pm
120$~pc.  Several other groups have estimated the distance to the
cluster using photometric parallax techniques and obtain results
ranging from $\approx$1100 to $\approx$ 1400~pc (Moffat \& Vogt 1973,
Fitzgerald 1987, Kalcheva \& Georgiev 1992).  Vazquez \& Feinstein
(1992) obtained an age of about 3 Myr by fitting the upper main
sequence, they also determined a distance of about 1410 $\pm 120$~pc
using an R~=~3.1 reddening law.  They noted that the steep reddening
found by Herbst \& Havlen (1977) was due to a binarity-induced color
shift.  Extinction is low (A$_V \sim 0.5$).

It has been difficult to obtain a good catalog of stellar members.
The 59 stars in Herbst \& Havlen (1977) are only identified on a star
chart.  Similar practices were used by Moffat \& Vogt (1973), Arnal et
al.~(1988) and Vazquez \& Feinstein (1992).  The deepest published
study of NGC 6193 is the $Chandra$ zero order image of the $Chandra$
gratings observation of HD 150135.  HD~150136 is a remarkable O+O
spectroscopic binary with a short 2.66 day orbital period (Niemela \&
Gamen 2005).  HD 150136 also has unusual radio properties, being a
non-thermal radio emitter (Benaglia, Cappa, \& Koribalski 2001) which
made it an interesting $Chandra$ gratings target in its own right
(Skinner et al.\ 2005).  This deep spectrum produces a useful
zero-order image.  The zero-order data include the central 2\arcmin
$\times$ 2\arcmin\ region of the cluster.
This observation, which should be fairly complete to 1 solar mass,
reveals 43 X-ray sources within 4 square arcminutes (Skinner et
al~2005).  Only 11 of these had previous optical identifications,
however all 43 were detected at H band and are likely cluster
members. In addition to those sources cataloged by Skinner et
al.~there are about 30 X-ray point sources visible outside of the
central 2\arcmin $\times$ 2\arcmin\ region in this data set. Most of
these have I-band magnitudes consistent with cluster membership and
are between 2\arcmin\ and 6.8\arcmin\ from HD 150136.  Wolk et al.\
(2008) obtained a 100 ks $Chandra$ observation of
RCW~108--IR.  Due to the orientation of the telescope pointing, about
64 square arcminutes of the NGC 6193 field was covered, mostly to the
south and west of HD 150136.  These data were a little less than 10
times deeper than the observation of Skinner et al.  An additional 99
X-ray sources associated with NGC 6193 were detected. Most of these
have infrared counterparts, which supports membership in Ara OB1a.
This brings the total identified membership of NGC~6193 up to about
200.  But we emphasize the non-uniformity of the data.  This is
especially true of the Chandra pointings which cover perhaps 20\% of
the area of NGC~6193.  A recent shallow $Spitzer/IRAC$ map of the
region detected about 185 Class~II sources, associated with both
NGC~6193 and RCW~108--IR, and 13 Class~I sources associated with
RCW~108--IR alone (Wolk et al.\ 2008).  But this observation was
centered on RCW~108--IR and so again does not provide a uniform sample
of NGC~6193.

\section{IRAS 16362--4845/RCW~108--IR}

In addition to the NGC 6193 cluster, the nearby source IRAS 16362--4845
harbors the extremely young stellar cluster RCW~108--IR, which has been
recently studied (Comer\'on et al.\ 2005, Comer\'on \& Schneider 2007,
Wolk et al.\ 2008). The bright rim NGC 6188 marks the border between
the HII region and a dense dark nebula containing RCW~108--IR.  The
bright rim is produced by the ionization of the molecular cloud
hosting IRAS 16362--4845 by the two central stars of NGC 6193.  The
dark cloud contains a high emission measure knot (Wilson et al. 1970),
which appears to be a compact HII region (Frogel \& Persson 1974).
The comprehensive UBVRI study by Herbst \& Havlen (1977) proposed this
was a site of very recent star formation.

\begin{figure}[!ht]
\begin{center}
\includegraphics[scale = 0.7, angle = 0]{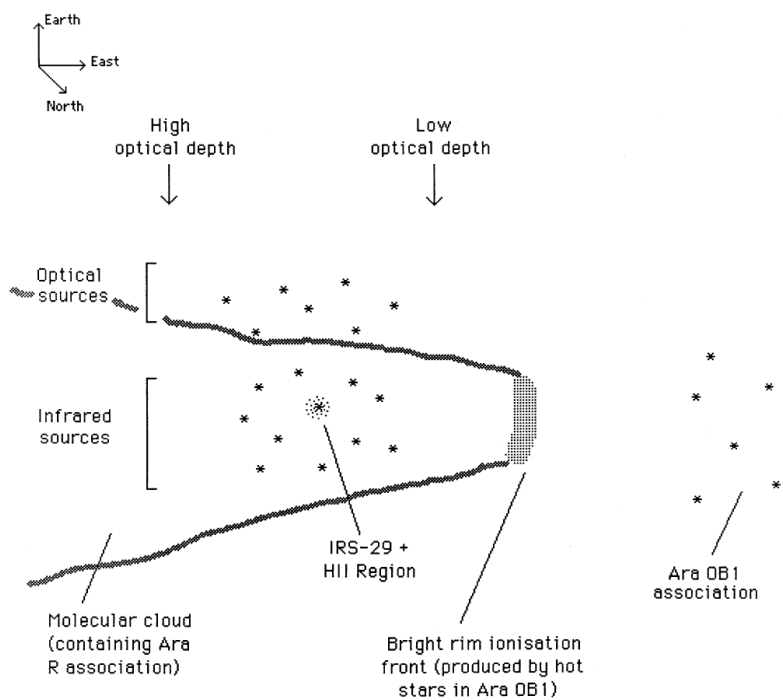}
\end{center}
\caption{ A simple model of the interface between RCW~108 and NGC~6193
from Straw et al. (1987).  It now appears that the older stars lie
somewhat behind the molecular cloud since none are seen superimposed
over it.} \label{straw}
\end{figure}

RCW~108--IR is a young, compact cluster partially embedded in its
parent molecular cloud.  Straw et al. (1987) used photometry at IR
wavelengths (1.2-100~$\mu$m) to perform the first spatially complete
survey of the cluster.  They report on 55 objects; $<$ 20 of these
have optical counterparts.  In addition to the point sources, there is
diffuse IR emission.  The total luminosity of the cluster is about
1.8$\times 10^5 $~L$_\odot$, dominated by at least two early O stars.
The full aggregate of O and B stars appears responsible for the
ionization of the diffuse emission (Comer\'on et al. 2005). The
primary exciting source star is identified as IRS~29 (Straw et al.\
1987) -- probably an O8 star.  Straw et al. also identify at least one
protostar. They put forward a physical model of the region in which
photoionization of the older stars in the Ara OB1a association ionized
the bright rim (NGC 6188) which lies at the edge of a finger of
molecular material which thickens from east to west
(Figure~\ref{straw}).

Several studies at radio wavelengths have been completed to understand
the physical properties of the dark cloud.  Goss \& Shaver (1970)
detected continuum emission at 5 GHz, Whiteoak \& Gardner (1974)
detected H$_2$CO absorption at 4830 MHz, and CO emission at 115 GHz
was detected by Whiteoak \& Otrupcek (1982).  The hydrogen
recombination lines H109$\alpha$ and H166$\alpha$ were also measured
(Wilson et al. 1970; Cersosimo 1982).  RCW~108 IRS~29 is the second
strongest water source seen by SWAS (second to the BN object in Orion;
Gary Melnick-private communication) with a peak flux of over 7000~Jy
in the 557~GHz line.  A follow up study suggests that gaseous H$_2$O
is largely restricted to a thin layer of gas near the cloud surface.

\begin{figure}[!ht]
\begin{center}
\plotfiddle{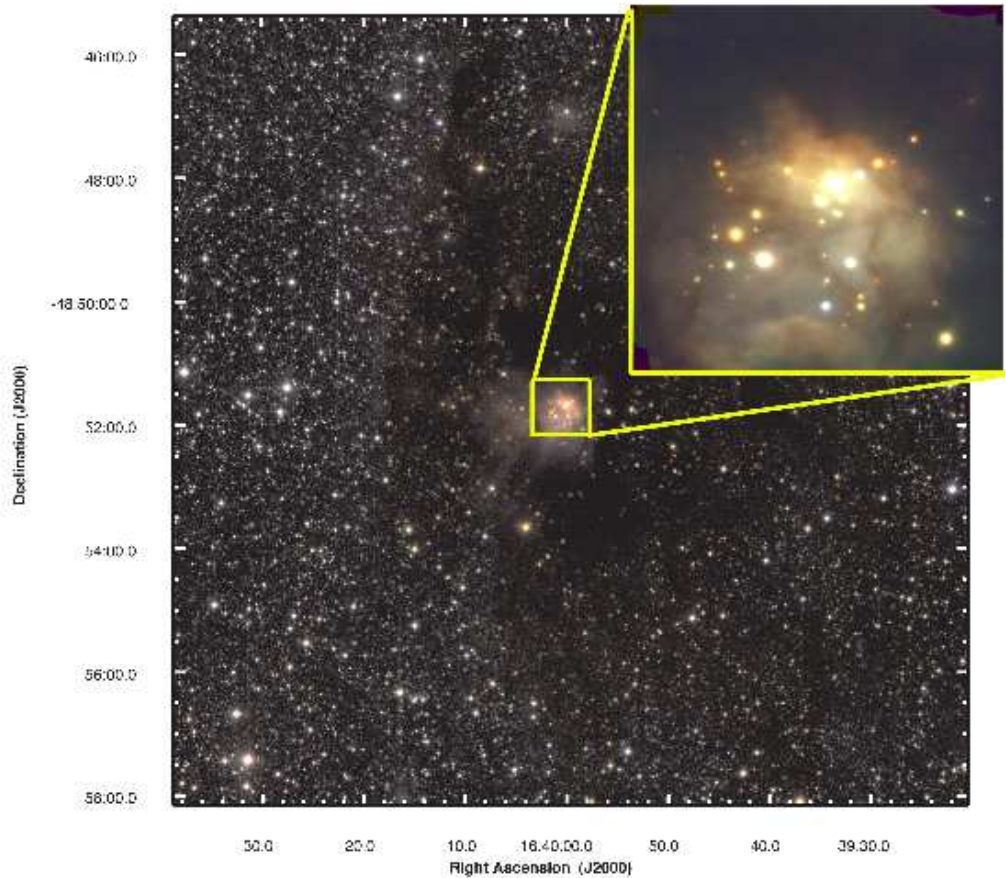}{11.0cm}{0.0}{72.0}{72.0}{-200.0}{-25.0}
\end{center}
\caption{A JHK$_S$ (blue, green and red respectively) image of
  RCW~108--IR from the NTT. The field of view is about 13\arcmin\ on a
  side, corresponding to about 6~pc on a side.  Five-sigma detection
  limits in the least-exposed areas of the image are typically J=18.6,
  H=17.7, K=17.2 mag. The cloud is very opaque immediately north and
  south of the emission nebula. Most of the stars with optically thick
  disks are either within this emission or along the eastern edge of
  the cloud. Inset: A J (blue), H (green), and K (red) color composite
  of RCW~108-IR obtained using adaptive optics at the VLT.  The frames
  are centered on the brightest source in the near-infrared, at
  RA(2000) = 16:40:00.2, Dec(2000) =-48:51:40. The field is 52\arcsec\
  $\times$ 57\arcsec.  }
\label{NTT}
\end{figure}

There have been two recent near-IR through mm studies of RCW~108--IR.
Urquhart et al. (2004) compared the 2MASS and MSX observations of the
cloud with radio recombination lines and radio continuum.  They find
that RCW~108--IR is an ultracompact HII region, less than 0.1~pc in scale
with a core--halo morphology.  They discuss 8 new sources within the
UCHII region and report 3 sources detected at 25~$\mu$m\ and set back
about 2.5\arcmin\ from the bright rim.  A more detailed study has been
recently published by Comer\'on et al. (2005).  From their high
resolution CO map they estimate a total mass of the cloud at 8000
M$_{\odot}$ with $<$ 200~M$_{\odot}$ in the molecular concentration
harboring the compact HII region. They produced a very high resolution
JHK$_s$ map of the region (Figure~\ref{NTT}). They identify 25 stars
whose luminosities suggest spectral types earlier than A0 under the
assumption that there is no significant circumstellar contribution to
the K-band flux and conclude that the ionization of the UCHII region is
provided by this aggregate.  They suggest that low--mass star
formation has yet to commence here. IRS 29 is found to be an O9 star
from the visible spectrum of the compact HII nebula, which is in
agreement with its infrared photometry.  Out of 4365 stars brighter
than K$_s$=14.5 in the whole 13\arcmin $\times$ 13\arcmin\ field, 87
are found to have strong disk signatures, most of them located in the
molecular cloud that contains RCW108--IR.

\begin{figure}[!ht]
\begin{center}
\includegraphics[scale = 0.7, angle = 0]{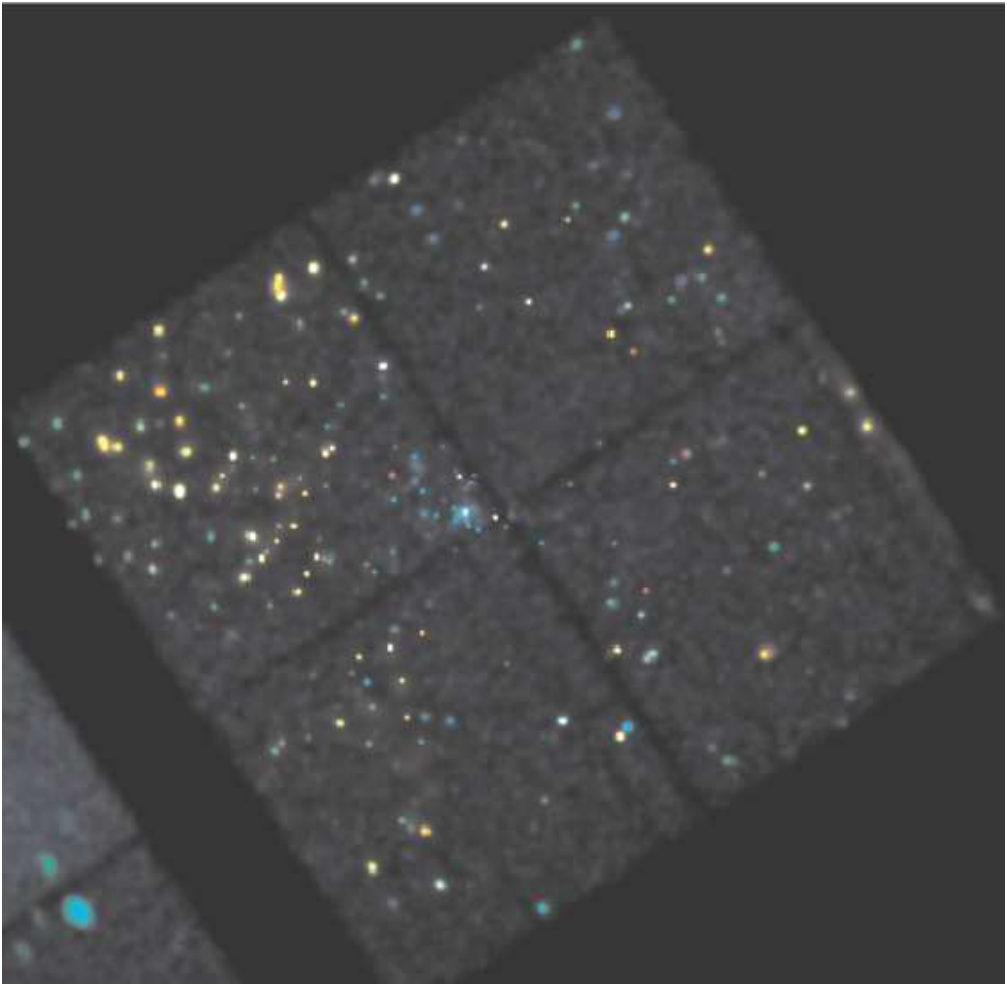}
\end{center}
\caption{$Chandra$/ACIS-I X-ray map of a somewhat wider region than
shown in Figure~\ref{NTT}.  The colors indicate
X-ray energy -- red for soft X-rays ($<1$ keV), blue for hard X-rays
($>$ 2.4 keV) and green in between. Hard X-rays penetrate gas better
than soft X-rays leading to X-ray "blueing" equivalent to
optical/infrared reddening.  Thus the stars associated with
RCW~108--IR appear blue and the stars to the east and south associated
with NGC~6193 appear yellow or red.  The region is about 20\arcmin\
across. From Wolk et al. (2008).} \label{Xray}
\end{figure}

In their 100 ks $Chandra$ pointing centered on RCW 108--IR, Wolk et
al.~(2008) detect 32 point sources in the central arcminute, and 65
within the central 2\arcmin (Figure~\ref{Xray}). They find a sharp
rise in the absorption column of log N$_H$ $>$ 21.2 (A$_V > 10$) for
these stars.  By contrast X-ray sources associated with NGC 6193 have
inferred extinctions between 0 and 5 A$_V$.  About 10 of the 65
sources in the core region do not have counterparts in the deep survey
of Comer\'on et al.

Recent JHK$_s$L$'$ observations of the stellar content of RCW~108-IR
have been carried out by Comer\'on \& Schneider (2007) using adaptive
optics-assisted imaging at the VLT. The superior resolution attained
with adaptive optics provides a much more detailed and deep view than
those available in previous studies. The deep imaging reveals faint
members of the cluster, probably including massive brown dwarfs. This
proves that, unlike suggested by Comer\'on et al.(2005), the cluster
does contain low-mass stars. However, such members are present in
numbers far below those expected when the bright end of the K-band
luminosity function is extrapolated to fainter magnitudes (see
e.g. Muench et al.  2002). This is interpreted as evidence for a
top-heavy initial mass function, perhaps related to special
circumstances that triggered star formation in the cluster (Sect.~6). The
revised mass of the cluster based on these new observations is
estimated to be $\sim 370$~M$_\odot$.  Comer\'on \& Schneider (2007)
also report the existence of a point-like Class~I source detected only
at 3.8~$\mu$m, as well as by {\it Spitzer}, projected on the border
between the HII region and its surrounding molecular cloud.

\section { NGC 6204 and NGC 6167}

NGC~6204 is an open cluster about 2 degrees northeast of NGC~6193.
Whiteoak (1963) identified 78 possible members. He noted a wide
variation in extinction, which prevented a good distance estimate.
Hogg (1965) identified this clump as two distinct clusters.  One
cluster is still known as NGC 6204 and is estimated to be at 1180 $\pm
50$~pc.  The stars in the other cluster, Hogg 22, are about twice as
distant and hence not part of Ara OB1a or Ara OB1b.  Forbes \& Short
(1996) performed photoelectric photometry on 36 member stars of NGC
6204 and find an age of about 125 Myr.  This is much older than the
other clusters in this region.  Thus, these stars are probably not
associated unless winds or shocks from extinct stars in this cluster
triggered all that followed.  The age difference between NGC 6204 and
NGC~6193/RCW108--IR seems to safely exclude any role of the cluster
in triggering star formation. Even if the velocity difference between
the cluster and NGC~6193 were just a fraction of the typical velocity
dispersion in a star forming complex, that would already imply that
NGC 6204 formed far from its current location and that it was benign
(with no stars with strong winds or able to explode as supernovae) by
the time it got close to NGC~6193 or RCW 108--IR.  Nonetheless,
Waldhausen et al. (1999) argue that NGC~6204 is associated with
NGC~6193 because their polarization angles are close (about 15$^{\rm
o}$).

NGC~6167 about 1 degree southwest of NGC~6193 was also poorly defined
in the original work by Whiteoak (1963) due to highly variable
absorption.  Br\"uck \& Smyth (1967) identified this as a real
cluster at 1200 $\pm 200$pc and 40 Myr old.  Moffat \& Vogt (1975)
determined a distance of 600 pc, however their argument against the
1.2 kpc distance rests on a single, perhaps foreground, star. Rizzo \&
Bajaja (1994) assert the age to be closer to 10 Myr.  Waldhausen et
al.~(1999) suggest that this is two overlain clusters since there
appear to be two polarizations (offset by 15$^{\rm o}$).

\section {IRAS~16379--4856 and  IRAS~16348--4849}

\begin{figure}[!t]
\begin{center}
\plotfiddle{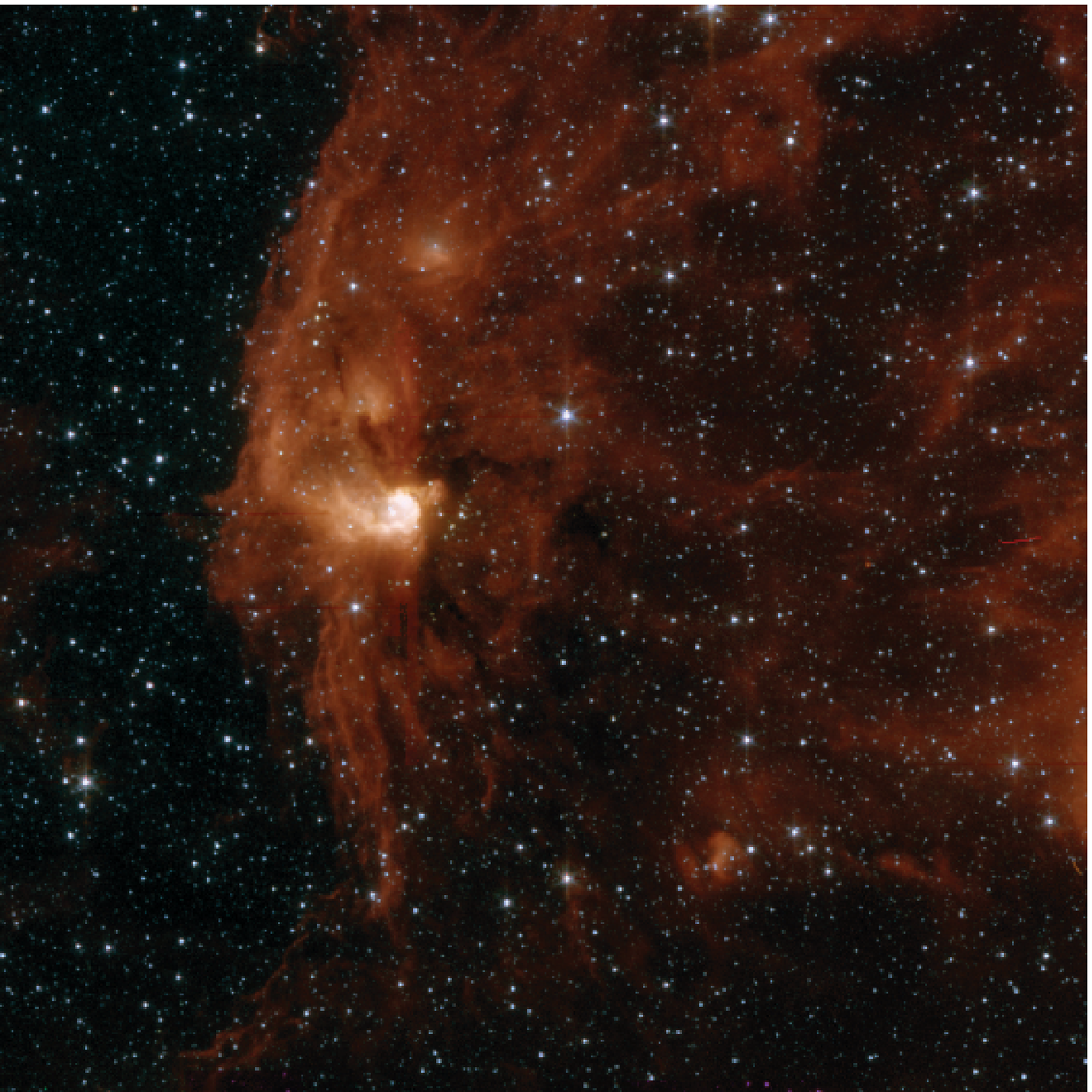}{12.5cm}{0.0}{60.0}{60.0}{-184.0}{-65.0}
\end{center}
\caption{$Spitzer$/IRAC image of RCW~108--IR (IRAC bands 1, 2, 3 and 4
are blue, yellow, orange and red respectively.)
Data are from a $Spitzer$ early release observation -- Melnick-PI.
The field of view is about 22\arcmin\ on a side, corresponding to
about 10~pc on a side.  The red is dominated by PAH emission that
fills much of the space and shows a deficit of stars in
Figure~\ref{NTT}. There are slight enhancements in the PAH emission on
the southeastern and western portions of the field associated with
IRAS~16379--4856 \& IRAS~16348--4849. Most of the stars with optically
thick disks are either within the core PAH emission or along the
eastern edge.}
\label{SPITZERfig}
\end{figure}

The two IRAS/CO peaks closest to RCW~108--IR are briefly discussed
here.  To the east is a small diffuse optical nebula near IRAS
16379--4856.  Wolk et al. (2008) detect a complex of bright unresolved
X--ray emission coincident with this region. A search of the 2MASS
catalog for K-band excess sources in this region reveals the strongest
concentration of such sources to be associated with this emission, at
least 25 sources with K-band excesses in a 2\arcmin\ $\times$ 1\arcmin\
region.  These sources all lie on the northwestern portion of a
circular ring of 8~$\mu$m emission about 10\arcmin\ in diameter.  There
are two 21~$\mu$m MSX point sources in the region.  One centered among
the 25 K-band excess sources, the other located near the inner edge of
the ring coincident with the centroid of the IRAS source.  This cloud
is coincident with region H of the CO survey by Arnal et al.
(2003). It is one of three CO clumps coincident with an IRAS source
(RCW 108--IR and IRAS~16348--4849 - see below -- are the other two).
They estimate the mass of the cloud to be 5800 M$_\odot$, second only
to the core region of RCW 108--IR.

MSX data indicate a mid-IR cloud about 10\arcmin\ west of RCW~108--IR
and coincident with IRAS~16348--4849 with a peak 8~$\mu$m flux of about
20\% that of RCW 108.  The overall extent of the cloud is about
10\arcmin\ on a side.  Tendrils of warm dust connecting the cloud and
RCW~108 are evident in the MSX and $Spitzer$ images. Two sources from
the recent $^{12}$CO observations by Arnal et al. (2005) lie within
this dark cloud.
Five B and A stars have been identified in this cloud; the most
massive is the B2IV star HD 149658. 2MASS data show about 20 stars in
the cloud with K$_S <$ 15 and K-band excesses indicative of disks.
Most of these stars lie within a circular region of 20~$\mu$m emission at
about 1\% the level of RCW 108--IR (Figure~\ref{SPITZERfig}).

These two regions, like RCW~108--IR, lie between NGC 6193 and the
putative center of expansion.  IRAS 16379--4856 is almost the same
projected distance as RCW 108 from HD 150135/6 at the center of
NGC~6193.  So, the nominal cluster near IRAS 16379--4856 appears to be
a sibling of RCW~108. Likewise formed by photoionization from the OB
stars in NGC 6193 striking a cloud which appears to have been swept
out by an event related to NGC~6167. The stars near IRAS~16357--4832
appear to compose a sibling cluster as well. The 8~$\mu$m morphology
suggests that it was torn off of RCW~108 itself.  This cluster is
intermediate in mass between the former two.  Additional bright CO and
8~$\mu$m emission is visible for a degree north of RCW 108.  While the
details as described here are speculative at this point, this is clearly a
dynamic and active association.

\section{Triggered Star Formation}

It is the morphology of the region that makes Ara OB1a
fascinating. NGC~6193 appears in relation to NGC 6188/IRAS 16362--4845
in much the same way that the $\sigma$ Orionis cluster (Walter et
al. 1997, see also chapter by Walter et al.) relates to the Horsehead
nebula.  The emission in the open cluster is radiation bounded by the
molecular cloud containing the embedded cluster RCW~108--IR, and is
spectacularly revealed in the figures as a clear ridge between them
(NGC~6188; Figure~\ref{ESO}).  This appears as one of the clearest
examples of triggered star formation in the sky.  But which mechanism
and how it works is still an open issue.

Herbst \& Havlen (1977) originally put forward the hypothesis that Ara
OB1a was an example of triggered star formation.  Their favored model
was that the trigger was a supernova event.  The $prima-facie$
evidence for this was the circular shape of the dust ring upon which
RCW 108--IR sits as a ``diamond.''  Straw et al. (1987) countered with
the fact that RCW~108--IR is surrounded by some more evolved objects
and point to the bright rim NGC~6188 as evidence that photoionization
is very strong in the region.  Thus, they favor the model originally
detailed by Elmegreen \& Lada (1977) in which successive sites of
star formation are triggered by photoionization from the previous
generation.

Radio observations have much to contribute to this portion of the
discussion since the radio data trace the gas and dust before the
onset of protostellar collapse. Phillips et al.~(1986) mapped the
region in CO and find a CO cloud centered 2--3\arcmin\ southeast of
the bright stars of NGC~6193.  They hypothesize that the cloud is
behind NGC~6193 and is being compressed by the winds of the OB
stars. However, there is no evidence for embedded sources within this
cloud. The fragmented structure of the cloud supports its
interpretation as the heavily eroded remnant of a now dispersed,
larger cloud, possibly the progenitor of NGC~6193 (Comer\'on et
al. 2005).

Arnal et al.~(1987) find a connection between the stellar population
of Ara OB1a and an expanding neutral hydrogen ring centered near
NGC~6167.  The connection is that they derive the same distance to the
gas as has been independently determined for the stars of $\approx
1400 pc$.  At that distance the HI ring is about 30~pc in radius and
20~pc in width expanding at 10~km~s$^{-1}$.  Thus, Arnal et al. derive
an upper age of this HI ring of 2$\times 10^6$ yr and an age closer to
1~Myr assuming a wind blown bubble (Weaver et al.~1977).  While this
is somewhat younger than current age estimates of NGC 6193, the age of
NGC~6193 is based on very few stars.  If the formation of NGC~6193
were the result of a supernova in NGC~6167 the problem of the age is
more severe and in the opposite sense as NGC~6193 should be quite
young, $\sim 0.1$ Myr, in this case (Straw et al. 1987).

\begin{figure}[!t]
\centering
\includegraphics[width=0.9\textwidth,draft=False]{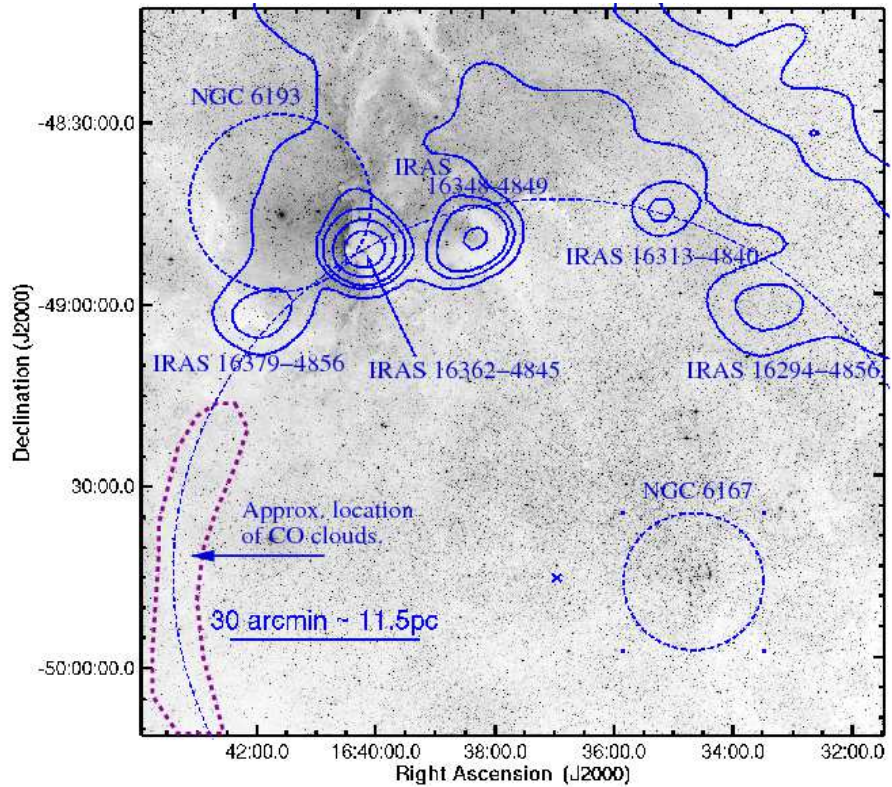}
\caption{A wide view of Ara OB1a.  The base image is a DSS2 red plate
about 2 degrees on a side. The top left quadrant is the region shown
in Figure 1. The clusters NGC~6193 and NGC~6167 are indicated as well
as RCW~108 and several IRAS point sources.  The three IRAS sources
nearest NGC 6193 all have 8~\micron\ counterparts in MSX data. IRAS
60~\micron\ contours of 300, 450, 600, 1200 and 2400 Jy/Sr are
indicated. The large circle is used to connect the various features.
There is a chain of CO clouds following along the eastern side of the
circle.} \label{IRAS}
\end{figure}

Rizzo \& Bajaja (1994) follow up on the Arnal et al.\ finding with a
more detailed HI study.  They confirm the initial result.  In their
model, the 10~Myr stars of NGC~6167 formed the HI shell. This
triggered the 3~Myr stars in NGC 6193.  In their model the HI shell
continues on to trigger RCW~108--IR.  However, RCW~108--IR is
misplaced in their Figure~6.  It seems more likely that
photoionization from the stars in NGC~6193 is triggering RCW~108 {\it
in the reverse direction} relative to the outward propagating HI ring.
Supporting evidence for external triggering in the clouds that host
RCW~108--IR is given by Comer\'on et al. (2005) who found the cluster
members they could identify are not symmetrically distributed
throughout the cloud, but clumped along the ionization front as well
as the core.

At a smaller scale, Comer\'on \& Schneider (2007) consider that their
findings on the stellar population of RCW~108-IR provide evidence that
its embedded cluster formed as a result of external triggering. In
particular, a top-heavy initial mass function like that derived by
those authors was suggested by Elmegreen \& Lada (1977) as a possible
byproduct of sequential star formation. This may be a consequence of
the warm temperature of the gas in the shocked layer located ahead of
the ionization front driven by the hot stars in NGC~6193, as well as
of turbulence induced by instabilities in that layer and of
coalescence of unstable fragments. Moreover, Comer\'on \& Schneider
(2007) also point out that the spatial distribution of massive stars
in RCW~108-IR differs from that commonly found in other clusters
containing high-mass members, which tend to concentrate towards the
cluster center as a result of the formation conditions and early
dynamical evolution (Bonnell et al.  2007). This is not expected for
externally triggered star formation, and indeed is not observed in
RCW~108-IR.

\section{The Future of Ara OB 1a}

Will more star formation take place in Ara OB1a?  We believe
this is the case and that it is already underway.  A recent CO survey of
the region finds over a dozen CO clouds in the region covering an area
almost 3 degrees on a side (Arnal et al. 2003).  While several groups
of clouds are kinematically coherent, there is a spread in velocities
of almost 20~km~s$^{-1}$.  Three of the concentrations are adjacent to
the bright rim separating RCW~108 and NGC~6193. IRAS data (see
Figure~\ref{IRAS}) show three concentrations of cool dust in the
vicinity of RCW~108--IR spaced about 5~pc apart. These concentrations
are bright from 8~$\mu$m to 100~$\mu$m.  The brightest of these is
RCW~108--IR itself. The IRAS peaks are coincident with $^{13}$CO peaks
found by Yamaguchi et al.~(1999). The $^{13}$CO peaks then make a
chain to the south. Arnal et al. (2003) identify five separate
$^{12}$CO features in this cloud. The chain extends almost 1.5$^{\rm
o}$ south of RCW~108--IR. In fact, the CO peaks taken in concert with
the 5 IRAS peaks (including IRAS 16379--4856 and IRAS 16348--4849
adjacent to RCW~108--IR) cover almost exactly 180 degrees of arc in a
near perfectly circular alignment 1.03$^{\rm o}$ from $l$=335:29:12
$b$=$-$1:42:23, about 8~pc from the center of NGC~6167.  More
specifically, Comer\'on et al. (2005) estimate a current star
formation efficiency in the main cloud hosting IRAS
16362--4845/RCW~108--IR to be well below 10\%, suggesting that most of
star formation in the cloud still has to take place. They also
speculate that its future evolution may lead to a complex similar to
the Orion Nebula Cluster, with the aggregate of IRAS 16362--4845 being
the equivalent to the Trapezium cluster.

\vspace{0.5cm}

{\bf Acknowledgments.}  We acknowledge many useful comments from the
anonymous referee, one of which was worked directly into the text.  We
gratefully acknowledge the financial support of NASA grant GO4-5013X
(Chandra) and from NASA contract NAS8-39073 (CXC). Archival data was
obtained from the Two Micron All Sky Survey (2MASS), a joint project
of the University of Massachusetts and the Infrared Processing and
Analysis Center/California Institute of Technology, funded by the
National Aeronautics and Space Administration and the National Science
Foundation and additional data were obtained with the Spitzer Space
Telescope, which is operated by the Jet Propulsion Laboratory,
California Institute of Technology under a contract with NASA.  This
research has made use of the SIMBAD database, operated at CDS,
Strasbourg, France, and of NASA's Astrophysics Data System
Bibliographic Services.



\begin{thebibliography}{}


\bibitem[Arnal et al.(1987)]{1987A&A...174...78A} Arnal, E.~M., Cersosimo, J.~C., May, J., \& Bronfman, L.\ 1987, \aap, 174, 78
\bibitem[Arnal et al.(2003)]{2003A&A...412..431A} Arnal, E.~M., May, J., \& Romero, G.~A.\ 2003, \aap, 412, 431
\bibitem[Arnal et al.(2005)]{2005ASPC..344..173A} Arnal, E.~M.,
  Romero, G.~A., May, J., \& Minniti, D.\ 2005, ASP Conf.~Ser.~344:
  {\em The Cool Universe: Observing Cosmic Dawn}, ed. C. Lidman \&
  D. Alloin, 173
\bibitem[Bonnell et al.(2007)]{2005PPV.344..173A} Bonnell, I.A., Larson, R.B., \& Zinnecker, H., 2007, in {\em Protostars and Planets V}, eds. B. Reipurth, D. Jewitt, K. Keil, Univ. of Arizona Press, 149
\bibitem[Benaglia et al.(2001)]{2001A&A...372..952B} Benaglia, P., Cappa,
C.~E., \& Koribalski, B.~S.\ 2001, \aap, 372, 952
\bibitem[Bruck \& Smyth(1967)]{1967MNRAS.136..431B} Br\"uck, M.~T. \& Smyth, M.~J.\ 1967, \mnras, 136, 431
\bibitem[Cersosimo(1982)]{1982ApL....22..157C} Cersosimo, J.~C.\ 1982, \apj, 22, L157
\bibitem[Comer{\'o}n et al.(2005)]{2005A&A...433..955C} Comer{\'o}n, F., Schneider, N., \& Russeil, D.\ 2005, \aap, 433, 955
\bibitem[Comer{\'o}n et al.(2007)]{2007A&A...433..955C} Comer\'on, F. \& Schneider, N., 2007, A\&A, 2007, 473, 149
\bibitem[Elmegreen \& Lada(1977)]{1977ApJ...214..725E} Elmegreen, B.~G. \& Lada, C.~J.\ 1977, \apj, 214, 725
\bibitem[Fitzgerald(1987)]{1987MNRAS.229..227F} Fitzgerald, M.~P.\ 1987, \mnras, 229, 227
\bibitem[Forbes \& Short(1996)]{1996AJ....111.1609F} Forbes, D. \& Short, S.\ 1996, \aj, 111, 1609
\bibitem[Frogel \& Persson(1974)]{1974ApJ...192..351F} Frogel, J.~A. \& Persson, S.~E.\ 1974, \apj, 192, 351
\bibitem[Herbst \& Havlen(1977)]{1977A&AS...30..279H} Herbst, W. \& Havlen, R.~J.\ 1977, \aaps, 30, 279
\bibitem[Herbst(1975)]{1975AJ.....80..498H} Herbst, W.\ 1975, \aj, 80, 498
\bibitem[Hogg(1965)]{1965PASP...77..440H} Hogg, A.~R.\ 1965, \pasp, 77, 440
\bibitem[Kaltcheva \& Georgiev(1992)]{1992MNRAS.259..166K} Kaltcheva, N.~T. \& Georgiev, L.~N.\ 1992, \mnras, 259, 166
\bibitem[Moffat \& Vogt(1973)]{1973A&AS...10..135M} Moffat, A.~F.~J. \& Vogt, N.\ 1973, \aaps, 10, 135
\bibitem[Moffat \& Vogt(1975)]{1975A&AS...20..155M} Moffat, A.~F.~J. \& Vogt, N.\ 1975, \aaps, 20, 155
\bibitem[Muench et al. (2002)]{2002APJ...20..155M}Muench, A.A., Lada, E.A., Lada, C.J., Alves, J., 2002, \apj, 573, 366
\bibitem[Niemela \& Gamen(2005)]{2005MNRAS.356..974N} Niemela, V.~S. \&
Gamen, R.~C.\ 2005, \mnras, 356, 974
\bibitem[Phillips et al.(1986)]{1986A&AS...65..465P} Phillips, J.~P., de Vries, C.~P., \& de Graauw, T.\ 1986, \aaps, 65, 465
\bibitem[Rizzo \& Bajaja(1994)]{1994A&A...289..922R} Rizzo, J.~R. \& Bajaja, E.\ 1994, \aap, 289, 922
\bibitem[Rodgers et al.(1960)]{1960MNRAS.121..103R} Rodgers, A.~W., Campbell, C.~T., \& Whiteoak, J.~B.\ 1960, \mnras, 121, 103
\bibitem[Shaver \& Goss(1970)]{1970AuJPA..14..133S} Shaver, P.~A. \& Goss, W.~M.\ 1970, Australian Journal of Physics Astrophysical Supplement, 14, 133
\bibitem[Skinner et al.(2005)]{2005MNRAS.361..191S} Skinner, S.~L., Zhekov, S.~A., Palla, F., \& Barbosa, C.~L.~D.~R.\ 2005, \mnras, 361, 191
\bibitem[Straw et al.(1987)]{1987ApJ...314..283S} Straw, S., Hyland, A.~R., Jones, T.~J., Harvey, P.~M., Wilking, B.~A., \& Joy, M.\ 1987, \apj, 314, 283
\bibitem[Urquhart et al.(2004)]{2004A&A...428..723U} Urquhart, J.~S., Thompson, M.~A., Morgan, L.~K., \& White, G.~J.\ 2004, \aap, 428, 723
\bibitem[Vazquez \& Feinstein(1992)]{1992A&AS...92..863V} Vazquez, R.~A. \& Feinstein, A.\ 1992, \aaps, 92, 863
\bibitem[Waldhausen et al.(1999)]{1999AJ....117.2882W} Waldhausen, S., Mart{\'{\i}}nez, R.~E., \& Feinstein, C.\ 1999, \aj, 117, 2882
\bibitem[Walter et al.(1997)]{1997MmSAI..68.1081W} Walter, F.~M., Wolk,S.~J., Freyberg, M., \& Schmitt, J.~H.~M.~M.\ 1997, Memorie della Societa Astronomica Italiana, 68, 1081
\bibitem[Weaver et al.(1977)]{1977ApJ...218..377W} Weaver, R., McCray, R., Castor, J., Shapiro, P., \& Moore, R.\ 1977, \apj, 218, 377
\bibitem[Whiteoak et al.(1982)]{1982PASAu...4..434W} Whiteoak, J.~B., Otrupcek, R.~E., \& Rennie, C.~J.\ 1982, Proceedings of the Astronomical Society of Australia, 4, 434
\bibitem[Whiteoak \& Gardner(1974)]{1974A&A....37..389W} Whiteoak, J.~B. \& Gardner, F.~F.\ 1974, \aap, 37, 389
\bibitem[Whiteoak(1963)]{1963MNRAS.125..105W} Whiteoak, J.~B.\ 1963, \mnras, 125, 105
\bibitem[Wilson et al.(1970)]{1970A&A.....6..364W} Wilson, T.~L., Mezger, P.~G., Gardner, F.~F., \& Milne, D.~K.\ 1970, \aap, 6, 364
\bibitem[n(0)]{} Wolk, S.~J., Spitzbart, B.~D., Bradley, D., Bourke, T.~L., Gutermuth, R.~A., Vigil, M., \& Comer\'on, F. 2008, AJ, 135, 693
\bibitem[Yamaguchi et al.(1999)]{1999PASJ...51..791Y} Yamaguchi, R., Saito, H., Mizuno, N., Mine, Y., Mizuno, A., Ogawa, H., \& Fukui, Y.\ 1999, \pasj, 51, 791


\end{thebibliography}
\end{document}